\documentclass[12pt,amsbsy]{article}
\usepackage{epsfig}

\newcommand{\beq}{\begin{equation}}
\newcommand{\eeq}{\end{equation}}
\newcommand{\beqn}{\begin{eqnarray}}
\newcommand{\eeqn}{\end{eqnarray}}

\newcommand{\B}{\mbox{${\mathbf B}$}}
\newcommand{\E}{\mbox{${\mathbf E}$}}

\newcommand{\n}{\mbox{${\mathbf n}$}}
\newcommand{\p}{\mbox{${\mathbf p}$}}

\newcommand{\s}{\mbox{${\mathbf s}$}}

\newcommand{\bL}{\mbox{${\mathbf L}$}}

\newcommand{\br}{\mbox{${\mathbf r}$}}
\newcommand{\bS}{\mbox{${\mathbf S}$}}

\newcommand{\bv}{\mbox{${\mathbf v}$}}

\newcommand{\x}{\mbox{${\mathbf x}$}}

\newcommand{\vom}{\mbox{{\boldmath$\omega$}}}

\newcommand{\Na}{\mbox{{\boldmath$\nabla$}}}
\newcommand{\si}{\mbox{{\boldmath$\sigma$}}}

\newcommand{\val}{\mbox{{\boldmath$\alpha$}}}

\newcommand{\al}{\mbox{${\alpha}$}}

\newcommand{\ga}{\mbox{${\gamma}$}}
\newcommand{\Ga}{\mbox{${\Gamma}$}}
\newcommand{\de}{\mbox{${\delta}$}}

\newcommand{\ka}{\mbox{${\kappa}$}}

\newcommand{\ep}{\mbox{${\varepsilon}$}}
\newcommand{\om}{\mbox{${\omega}$}}

\newcommand{\pa}{\mbox{${\partial}$}}

\begin{document}

\begin{center}
{\large\bf  Spinning Relativistic Particles in External Fields}
\end{center}

\begin{center}
I.B. Khriplovich\footnote{khriplovich@inp.nsk.su}\\
\end{center}

\begin{center}
Budker Institute of Nuclear Physics, 630090 Novosibirsk, Russia,
\protect\newline and Novosibirsk University
\end{center}

\begin{abstract}
The motion of spinning relativistic particles in external
electromagnetic and gravitational fields is considered. A simple
derivation of the spin interaction with gravitational field
is presented. The self-consistent
description of the spin corrections to the equations of motion is
built with the noncovariant description of spin and with the
usual, ``na\"{\i}ve'' definition of the coordinate of a
relativistic particle.
\end{abstract}

\section{Introduction}
The pioneering paper~\cite{mm} by Myron Mathisson on the
relativistic equations of motion for spinning particle in a
gravitational field was published in Acta Physica Polonica. I
truly appreciate the invitation to submit a review on the subject
to this journal.

The general problem of the motion of a relativistic particle with
internal angular momentum (spin) in external electromagnetic and
gravitational fields consists of two parts: the description of the
spin precession and accounting for the spin influence on the
trajectory of particle motion.

As to the first part of the problem, that of the spin evolution by
itself, it is essentially settled, starting with the paper by
Mathisson \cite{mm} (see also \cite{pa}). A relatively simple
derivation of the corresponding relativistic equations, both in
electrodynamics and gravity, is presented below, together with the
detailed discussion of the limits of applicability for these
equations.

The situation is different with the second part of the problem,
i.e., how the interaction of spin with an external field
influences the trajectory of a particle. On the one hand, there is
an old prejudice according to which for elementary particles this
influence is unobservable by virtue of the uncertainty relation.
On the other hand, there are serious disagreements on the exact
form of the effect.

The discussed effect in electromagnetic field is of real physical interest,
being related to the problem of separating different
polarizations of relativistic particles in accelerators. However,
as mentioned, it is not as obvious whether one can observe in
practice the discussed spin corrections to the equations of motion
of elementary particles, for instance, electron or proton.
According to the well-known observation by Bohr (quoted
in~\cite{pau}), an additional Lorentz force due to the finite size
of the wave packet of a charged particle and to the uncertainty
relation, is on the same order of magnitude as the corresponding
component of the Stern-Gerlach force. However, this argument by
itself does not exclude in principle the possibility to observe a
regular Stern-Gerlach effect, even a small one, in the presence of
a comparable background due to the uncertainty relation. It is
sufficient to recall quite common situation when the accuracy with
which the energy of an unstable level is known, is much better
than the width of this level.

Besides, not only spin-dependent correlations certainly exist in
differential cross sections of scattering processes, but they are
effectively used to separate particles of different polarizations.
That is why it was proposed long ago to separate
charged particles of different polarizations in a storage ring through the spin
interaction with external fields~\cite{ros}. Though this proposal
is discussed rather actively (see, for instance, review~\cite{hei}), it is
not clear up to now whether it is feasible technically.

On the other hand, certainly there are macroscopic objects for
which internal rotation influences their trajectories. We mean the
motion of Kerr black holes in external gravitational fields. This
problem is of importance in particular for the calculation of the
gravitational radiation of binary stars. In this connection it was
considered in~[6 -- 9]. However, the equations of motion taking
account of the spin influence to lowest nonvanishing order in
$c^{-2}$, used in these papers, lead to results which differ from
the well-known gravitational spin-orbit interaction even in the
simpler case of an external field. As will be demonstrated below,
the disagreement originates from different definitions of the
center-of-mass coordinate. In the analysis of this disagreement
and in the solution of the problem we follow~\cite{khp, khp1}.

\section{Equations of Motion of Spin in Electromagnetic Field}

The equations of motion for spin of a relativistic particle in
electromagnetic field are well known. However, since the general
problem of the spin interaction with external gravitational field
reduces to the analogous problem for the case of an external
electromagnetic field, it is pertinent to start the detailed
analysis of the subject in this way. Besides, the subtleties with
the definition of the center-of-mass coordinate can be essentially
elucidated in this problem, which is somewhat more simple than the
gravitational one.

\subsection{Covariant Equation of Motion of Spin}

Let us consider at first the spin precession for a nonrelativistic
charged particle. The equation that describes this precession is
well known:
\beq\label{sB}
\dot{{\bf s}}= \,\frac{eg}{2m}\,[{\bf s} \times \B]\,.
\eeq
Here $\B$ is an external magnetic field, $e$ and $m$ are the
charge and mass of the particle, $g$ is its gyromagnetic ratio
(for electron $g\approx 2$). In other words, the spin precesses
around the direction of magnetic field with the frequency
$-(eg/2m)\B$. In the same nonrelativistic limit the velocity
precesses around the direction of $\B$ with the frequency $-
(e/m)\B$:
\beq\label{vB}
\dot{\bv}= \,\frac{e}{m}\,[\bv \times \B]\,.
\eeq
Thus, for $g=2$ spin and velocity precess with the same frequency,
so that the angle between them is conserved.

Let us note that both equations, (\ref{sB}) and (\ref{vB}), hold
as Heisenberg equations of motion in an external field for the
spin and velocity operators, $\s$ and $\bv$. On the other hand,
being averaged over properly localized wave packets, these
equations go over into the (semi)classical equations of motion for
spin and velocity. This refers also to the relativistic
generalizations of equation (\ref{sB}), discussed in this section
below.

We will consider at first the covariant semiclassical formalism
using the four-dimensional vector of spin $S_\mu$. This 4-vector
is defined as follows. In the particle rest frame $S_\mu$ has no
time component and reduces to the common three-dimensional vector
of spin ${\bf s}$, i.e., in this frame $S_\mu=(0,{\bf s})$. In the
reference frame where the particle moves with velocity $\bv$, the
vector $S_\mu$ is constructed from $(0,{\bf s})$ by means of the
Lorentz transformation, so that here
\beq\label{sS}
S_0=\ga\bv{\bf s}\,, \quad \bS ={\bf s}+\,\frac{\ga^2\bv(\bv{\bf
s})}{\ga+1}\,.
\eeq
Then, just by definition of $S_\mu$, the following identities take
place:
\beq\label{id}
S_\mu S_\mu = - {\bf s}^2\,(=\mbox{const}), \quad S_\mu u_\mu=0\,;
\eeq
as usual, here $u_\mu$ is the four-velocity. It should be
emphasized that it is just the velocity, which is a parameter of
the Lorentz group, that enters these relations (but not the
canonical momentum, which is not even a gauge-invariant vector).

The right-hand side of the equation for $dS_\mu/d\tau$ (here and
below $\tau$ is the proper time) should be linear and homogeneous
both in the electromagnetic field strength $F_{\mu\nu}$, and in
the same four-vector $S_\mu$, and may depend also on $u_\mu$. By
virtue of the first identity (\ref{id}), the right-hand side
should be four-dimensionally orthogonal to $S_\mu$. Therefore, the
general structure of the equation we are looking for is
\beq\label{bmt0}
\frac{dS_\mu}{d\tau}\,=\al F_{\mu\nu}S_\nu + \beta u_\mu
F_{\nu\lambda} u_\nu S_\lambda\,.
\eeq
Comparing the nonrelativistic limit of this equation with
(\ref{sB}), we find
\[
\al= \,{eg \over 2m}\,.
\]
Now we take into account the second identity (\ref{id}), which
after differentiation in $\tau$ gives
\[
u_\mu\,\frac{dS_\mu}{d\tau}\,=-S_\mu\,\frac{du_\mu}{d\tau}\,,
\]
and recall the classical equation of motion for a charge:
\beq\label{lor}
m\,\frac{du_\mu}{d\tau}\,= e F_{\mu\nu}u_\nu\,.
\eeq
Then, multiplying equation (\ref{bmt0}) by $u_\mu$, we obtain
\[
\beta = -\,\frac{e}{2m}\,(g-2)\,.
\]
Thus, the covariant equation of motion for spin is~[12 -- 14]
\beq\label{bmt}
\frac{dS_\mu}{d\tau}\,=\,\frac{eg}{2m} F_{\mu\nu}S_\nu
-\,\frac{e}{2m}\,(g-2) u_\mu F_{\nu\lambda} u_\nu S_\lambda\,.
\eeq

Let us discuss the limits of applicability for this equation.

Of course, typical distances at which the trajectory changes (for
instance, the Larmor radius in a magnetic field) should be large
as compared to the de~Broglie wave length $\hbar/p$ of the
elementary particle. Then, the external field itself should not
change essentially at the distances on the order of both the
de~Broglie wave length $\hbar/p$ and the Compton wave length
$\hbar/(mc)$ of the particle. In particular, if the last condition
does not hold, the scatter of velocities in the rest frame is not
small as compared to $c$, and one cannot use in this frame the
nonrelativistic formulae.

Besides, if the external field changes rapidly, the motion of spin
will be influenced by interaction of higher electromagnetic
multipoles of the particle with field gradients. For a particle of
spin $1/2$ higher multipoles are absent, and the
gradient-dependent effects are due to finite form factors of the
particle. These effects start here at least in second order in
field gradients and usually are negligible.

At last, in equation (\ref{bmt}) we confine to effects of first
order in the external field. This approximation relies in fact on
the implicit assumption that the first-order interaction with the
external field is less than the excitation energy of the spinning
system. Usually this assumption is true and the first-order
equation (\ref{bmt}) is valid. Still, one can easily point out
situations when this is not the case. To be definite, let us
consider the hydrogen-like ion $^3{\rm He}^+$ in the ground
$s$-state with the total spin $F = 1$. It can be easily
demonstrated that an already quite moderate external magnetic
field is sufficient to break the hyperfine interaction between the
electron and nuclear magnetic moments (a sort of Paschen-Back
effect). Then, instead of a precession of the total spin ${\bf F}$
of the ion, which should be described by equations (\ref{sB}) or
(\ref{bmt}) with a corresponding ion $g$-factor, we will have a
separate precession of the decoupled electron and nuclear spins.

Let us go back now to equation (\ref{bmt}). We note that for $g=2$
and in the absence of electric field, its zeroth component reduces
to
\[
\frac{dS_0}{d\tau}\,=0.
\]
Taking into account definition (\ref{sS}) for $S_0$ and the fact
that in a magnetic field a particle energy remains constant, we
find immediately that the projection of spin ${\bf s}$ onto
velocity, so-called helicity, is conserved.

\subsection{Noncovariant Equation of Motion for Spin\\
of Relativistic Particle. Thomas Precession}

We will obtain now the relativistic equation for the
three-dimensional vector of spin ${\bf s}$, that directly
describes the internal angular momentum of a particle in its
``momentary'' rest frame. This equation can be derived from
(\ref{bmt}) using relations (\ref{sS}), together with the
equations of motion for a charge in external field. It will
require, however, quite tedious calculations. Therefore, we choose
another way, somewhat more simple and much more instructive.

First, we transform equation (\ref{sB}) from the comoving inertial
frame, where the particle is at rest, into the laboratory one. The
magnetic field ${\bf B}^{\prime}$ in the rest frame is expressed
via the electric and magnetic fields ${\bf E}$ and ${\bf B}$ given
in the laboratory frame, as follows:
\[
{\bf B}^{\prime}=\gamma {\bf B} -\,\frac{\gamma^2}{\gamma +
1}\,{\bf v}({\bf v} {\bf B})- \gamma {\bf v} \times {\bf E}\,.
\]
This expression can be easily checked by comparing it component by
component with the transformation of magnetic field for two cases:
when this field is parallel to the velocity and orthogonal to it,
respectively. Then one should take into account that the frequency
in the laboratory time $t$ is $\gamma$ times smaller than the
frequency in the laboratory time $\tau$ (indeed, $d/dt =
d\tau/dt\cdot d/d\tau =\gamma^{-1}d/d\tau$). Found in this way
contribution to the precession frequency is
\[
\vom_g = -\,\frac{eg}{2m}\left[{\bf B} -\,\frac{\gamma}{\gamma +
1}\,{\bf v}({\bf v} {\bf B})- {\bf v} \times {\bf E} \right].
\]

However it is clear from equation (\ref{bmt}) that spin precesses
even if $g=0$. To elucidate the origin of this effect, the
so-called Thomas precession \cite{th1}, we consider two successive
Lorentz transformations: at first from the laboratory frame $S$
into the frame $S^\prime$ that moves with the velocity $\bv$ with
respect to $S$, and then from $S^\prime$ into the frame $S^{\prime
\prime}$ that moves with respect to $S^\prime$ with the
infinitesimal velocity $d\bv$. Let us recall in this connection
the following fact related to usual three-dimensional rotations:
the result of two successive rotations with respect to
noncollinear axes $\n_1$ and $\n_2$ contains in particular a
rotation around the axis directed along their vector product $\n_1
\times \n_2$. Now it is only natural to assume that the result of
the above successive Lorentz transformations will contain in
particular a usual rotation around the axis directed along $d\bv
\times \bv$. In result, spin in the rest frame will rotate in the
opposite direction by an angle which we denote by $\kappa \, [\,
d\bv \times \bv\, ]$. Here $\kappa$ is some numerical factor to
be determined below. It depends generally speaking on the particle
energy.

This is in fact the Thomas precession. Its frequency in the proper
time $\tau$ is
\[
\vom_T^\prime = \kappa [\, d\bv/d\tau \times \bv\, ] = \ka\,{e
\over m}\, [\,{\bf E}^\prime \times \bv\, ].
\]
Now we transform the electric field ${\bf E}^\prime$ from the
proper frame into the laboratory one, as it was done above for the
magnetic field ${\bf B}^{\prime}$, and go over also from the
proper time $\tau$ to $t$. In result, the frequency of the Thomas
precession in the laboratory frame is
\[
\vom_T = \kappa\,{e \over m}\,\left[\left ({\bf
E}-\,\frac{\ga}{\ga +1}\,\bv(\bv{\bf E})+ \bv \times {\bf
B}\right) \times \bv\right ]
\]
\[
= - \kappa\,{e \over m}\,\left[\bv \times {\bf E} -v^2 {\bf B}
+ \bv(\bv{\bf B}) \right ]\,. \;\quad\quad
\]
To find the coefficient $\kappa$, we recall that in a magnetic
field, for $g=2$ the projection of spin onto the velocity is
conserved. In other words, in this case the total frequency of the
spin precession $\vom = \vom_g + \vom_T$ coincides with the
frequency of the velocity precession which is well known to be
\[
\vom_v = -\,\frac{e}{m \ga}\,{\bf B}\,.
\]
From this we find easily that $\kappa= \ga/(\ga +1)\,$.
Correspondingly, the relativistic equation of motion for the
three-dimensional vector of spin ${\bf s}$ in external
electromagnetic field is
\[
\frac{d{\bf s}}{dt}\,= (\vom_g +\vom_T)\times {\bf s}
=\,\frac{e}{2m}\,\left\{\left(g-2+ \,\frac{2}{\gamma} \right)
[{\bf s}\times \B]\right.
\]
\beq\label{bmt1}
- \,(g-2) \,\frac{\gamma}{\gamma+1}\,[{\bf s}\times\bv](\bv\B)
\left. -\left(g-\,\frac{2\gamma}{\gamma+1}\right)\bigl[{\bf
s}\times[\bv \times \mathbf{E}]\bigr]\right\}.
\eeq

\subsection{Relativistic Spin Hamiltonian}

The relativistic Hamiltonian for the interaction of the
three-dimensional vector of spin with external electromagnetic
field is written in the usual form:
\beq\label{reha}
H = \vom {\bf s}\,.
\eeq
Not only does it generate via the standard relation
\beq
\frac{d\bf s}{dt}\,= \,\frac{i}{\hbar}\, [H, {\bf s}]
\eeq
equation (\ref{bmt1}). For instance, it is easy to
obtain with this Hamiltonian equations of motion of the quadrupole
moment of a relativistic particle in electric and magnetic fields,
neglecting the field gradients. In the particle rest frame, the
operator of its quadrupole moment is
\[
q_{mn}=\frac{3q}{2s(2s-1)}\left[s_m s_n + s_n s_m -
\frac{2}{3}\,s(s+1)\de_{mn}\right];
\]
here the structure in square brackets guarantees the symmetry and
vanishing trace of this operator, $q_{mn}=q_{nm}\,,\; q_{mm}=0$;
the overall factor at the square brackets corresponds to the
normalization condition $q_{zz} = q$ for $s_z = s$. To calculate
the commutator in the corresponding equation
\beq
\frac{d q_{mn}}{dt}\,= \,\frac{i}{\hbar}\, [\om_k s_k, q_{mn}]\,,
\eeq
is an elementary problem.

Moreover, Hamiltonian (\ref{reha}) is effective for the solution
of various other physical problems, from the single-photon
radiative transition between atomic $s$-levels to the low-energy
theorems for Compton scattering. So, the validity of this Hamiltonian
is beyond any doubts.

In particular, interaction (\ref{reha}) can be effectively used to
derive the additional spin-dependent force acting upon a charged
particle in an external electromagnetic field~\cite{dk}. However, since
$\vom$ depends directly on the particle velocity $\bv$ (but not on its
momentum $\p$), it is somewhat more convenient to employ here the Lagrangian
(but not Hamiltonian) formalism with
\beq\label{la}
L = - \,\vom \s\,.
\eeq
We will come back to the spin-dependent forces later.

\section{Spin Precession in Gravitational Field}
In this section we present a simple and general derivation of the
equations of the spin precession in a gravitational field
(restricting to first order in spin), based on a remarkable
analogy between gravitational and electromagnetic fields. Due to
this correspondence, the formulae of the previous section are
naturally adapted for the case of an external gravitational field.
In this way we easily reproduce and generalize known results for
gravitational spin effects.

\subsection{General Relations}
It follows from the angular momentum conservation in flat
space-time taken together with the equivalence principle that the
4-vector of spin $S^\mu$ is parallel transported along the
particle world-line. The parallel transport of a vector along a
geodesic $x^\mu(\tau)$ means that its covariant derivative
vanishes:
\begin{equation}\label{par}
\frac{DS^\mu}{D\tau}=\,0\,.
\end{equation}
We will use the tetrad formalism natural for the description of
spin. The tetrad components of spin
\[
S^a=\,S^\mu e^a_{\mu}
\]
(by the first letters of the Latin alphabet, $a,b,c,d,$ we label
here and below four-dimensional tetrad indices) behave as vectors
under Lorentz transformations of the locally inertial frame.
However, they do not change under generally covariant
transformations $x^\mu=f^\mu(x')$. In other words, the four
components $S^a$ are world scalars. Therefore, in virtue of
relation (\ref{par}), the equations for them appear as follows:
\begin{equation}\label{pars}
\frac{dS^a}{d\tau}=\,\frac{DS^a}{D\tau}=\,S^\mu
e^a_{\mu;\,\nu}u^\nu= \,\eta^{ab}\gamma_{bcd}\,u^d S^c\,.
\end{equation}
The covariant derivative of a tetrad is by definition
\[
e^a_{\mu;\,\nu}=\pa_\nu
e^a_{\mu}-\Ga^{\kappa}_{\mu\nu}e^a_{\kappa}\,,
\]
and the quantity
\beq
\gamma_{abc}=\,e_{a\mu;\,\nu}\,e^\mu_{b}e^\nu_{c}
\eeq
is the Ricci rotation coefficient. By means of covariant
differentiation of the identity $e_{a\mu}\,e^\mu_{b} = \eta_{ab}$,
one can easily demonstrate that these coefficients are
antisymmetric in the first pair of indices:
\beq\label{rota}
\gamma_{abc}=-\gamma_{bac}\,.
\eeq

Of course, the equations for the tetrad components of a 4-velocity
look exactly in the same way as those for spin:
\begin{equation}\label{paru}
\frac{du^a}{d\tau}=\,\eta^{ab}\gamma_{bcd}\,u^d u^c\,.
\end{equation}

The meaning of equations (\ref{pars}) and (\ref{paru}) is clear:
the tetrad components of both vectors vary in the same way since
their variation is due only to the rotation of the local Lorentz
frame.

There is a remarkable similarity between the discussed problem and
the special case of $g=2$ in electrodynamics. According to
equations (\ref{bmt}) and (\ref{lor}), the four-dimensional spin
and four-dimensional velocity of a charged particle with the
gyromagnetic ratio $g=2$ precess with the same angular velocity:
\[
\frac{dS_a}{d\tau}=\,\frac{e}{m}\,F_{ab}S^b, \quad
\frac{du_a}{d\tau}=\,\frac{e}{m}\,F_{ab}u^{b}.
\]
In other words, the obvious correspondence takes place:
\beq\label{corcov}
\frac{e}{m}\,F_{ab} \longleftrightarrow \gamma_{abc}u^c.
\eeq
It allows us to derive the precession frequency $\vom$ of a
three-dimensional vector of spin ${\bf s}$ in an external
gravitational field from expression (\ref{bmt1}) by means of the
simple substitution
\beq\label{cornon}
\frac{e}{m}\,B_i \longrightarrow
-\,\frac{1}{2}\,\epsilon_{ikl}\gamma_{klc}u^c; \;\;\;
\frac{e}{m}\,E_i \longrightarrow \gamma_{0ic}u^c
\eeq
(here again, it is just the velocity, a parameter of the local
Lorentz group, that enters these relations, but not the canonical
momentum). Thus, the precession frequency is
\begin{equation}\label{og1}
\omega_i=\,\epsilon_{ikl}\left(\frac{1}{2}\,\gamma_{klc}+
\,\frac{u^k}{u^0+1}\,\gamma_{0lc}\right)\frac{u^c}{u^0_w}\,.
\end{equation}
The factor $1/u^0_w$ in expression (\ref{og1}) is due to the
transition in the left-hand side of equation (\ref{pars}) to
differentiating over the world time $t$:
\[
\frac{d}{d\tau}=\,\frac{dt}{d\tau}\,\frac{d}{dt}
=\,u^0_w\,\frac{d}{dt}.
\]
We supply here $u^0_w$ with the subscript $w$ to indicate that
this is the world, but not the tetrad, component of 4-velocity.
All other indices in (\ref{og1}) are tetrad ones, $\;c=0,1,2,3,
\;\;i,k,l=1,2,3$.

Naturally, the relativistic Hamiltonian for the interaction of the
three-dimensional vector of spin with external gravitational field
is exactly of the same form~(\ref{reha}), as that in the
electromagnetic case. Here, of course, the precession frequency
$\vom$ is given by formula (\ref{og1}).

As to the limits of applicability of the presented equations,
which describe the spin precession in an external gravitational
field, they are quite analogous to those pointed out in subsection
2.1 for the case of an electromagnetic field.

However, in some respect the spin interaction with a gravitational
field differs essentially from that with an electromagnetic field.
In the case of an electromagnetic field, the interaction depends,
generally speaking, on a free phenomenological parameter,
$g$-factor. Moreover, if one allows for the violation of
invariance both under the reflection of space coordinates and
under time reversal, one more parameter arises in the case of
electromagnetic interaction, the value of the electric dipole
moment of the particle. The point is that both magnetic and
electric dipole moments interact with the electromagnetic field
strength, so that this interaction is gauge-invariant for any
value of these moments. Only the spin-independent interaction with
the electromagnetic vector potential is fixed by the charge
conservation and gauge invariance. On the contrary, the Ricci
rotation coefficients $\gamma_{abc}$ entering the gravitational
first-order spin interaction (\ref{pars}), as distinct from the
Riemann tensor, are noncovariant. Therefore, the discussed
interaction of spin with gravitational field is fixed in unique
way by the law of angular momentum conservation in flat space-time
taken together with the equivalence principle, and thus it
contains no free parameters~\cite{dau}.

On the other hand, it is no surprise that the precession frequency
$\vom$ depends not on the Riemann tensor, but on the rotation
coefficients. Of course, this frequency should not be a tensor: it
is sufficient to recall that a spin, which is at rest in an
inertial reference frame, precesses in a rotating one.

\subsection{Spin-Orbit Interaction. Weak Field}

One can check easily that in the weak-field approximation where
\[
g_{\mu\nu}=\eta_{\mu\nu}+ h_{\mu\nu},\quad |h_{\mu\nu}|\ll 1,
\]
there is no difference between the tetrad and world indices in
$e_{a\mu}$, and the tetrad appears as follows:
\[
e_{\mu\nu}=\eta_{\mu\nu}+ \tilde{e}_{\mu\nu},\quad
|\tilde{e}_{\mu\nu}|\ll 1.
\]
Relation between the tetrads and metric
\[
e_{a\mu}e_{b\nu}\eta^{ab}=g_{\mu\nu}
\]
in the weak-field approximation reduces to
\[
\tilde{e}_{\mu\nu}+\tilde{e}_{\nu\mu}=h_{\mu\nu}\,.
\]
Under the demand that tetrads are expressed via metric only, one
arrives at the so-called symmetric gauge for the tetrads where
\[
\tilde{e}_{\mu\nu}=\,{1 \over 2}\,h_{\mu\nu}\,.
\]
Then in the weak-field approximation the Ricci coefficients are:
\begin{equation}\label{gam}
\gamma_{abc}=\,\frac{1}{2}\,(h_{bc,\,a}-\,h_{ac,\,b})\,.
\end{equation}

Now, with relations (\ref{og1}) and (\ref{gam}) one can solve, for
instance, in an elementary way the problem of the gravitational
spin-orbit interaction for arbitrary particle velocities. In
the centrally symmetric field created by a mass $M$, the metric is
\begin{equation}\label{lin}
h_{00}=\,-\,{r_g \over r}\,=-\,\frac{2kM}{r}\,, \quad
h_{mn}=\,-\,{r_g \over
r}\,\delta_{mn}=\,-\frac{2kM}{r}\,\delta_{mn}\,.
\end{equation}
Here the nonvanishing Ricci coefficients are
\begin{equation}\label{ric}
\gamma_{ijk}=\,\frac{kM}{r^3}\,(\delta_{jk}r_i-\,\delta_{ik}r_j)\,,
\quad \gamma_{0i0}=\,-\,\frac{kM}{r^3}\,r_i\,.
\end{equation}
Plugging these expressions into formula (\ref{og1}) yields the
following result for the precession frequency:
\begin{equation}\label{so}
{\vom_{ls}}=\,\frac{2\gamma+\,1}{\gamma+\,1}\,\frac{kM}{r^3}\, \bv
\times \br \,.
\end{equation}
The combination of a high velocity for a spinning particle with a weak
gravitational field refers obviously to a scattering problem.
Another possible application is to a spinning particle bound by
other forces, for instance, by electromagnetic ones, when we are
looking for the correction to the precession frequency due to the
gravitational interaction.

In the limit of low velocities, $\gamma \rightarrow 1$, formula
(\ref{so}) goes over into the classical result \cite{fok}
\begin{equation}
{\vom_{ls}}=\,\frac{3}{2}\,\frac{kM}{r^3}\, \bv \times \br \,,
\end{equation}
with the gravitational spin-orbit potential \footnote{It is
curious that this result by Fokker for the gravitational
spin-orbit interaction preceded by 5 years the corresponding one
by Thomas for the electromagnetic case.}
\begin{equation}\label{uls}
U_{ls}(\br)=\,\frac{3}{2}\,\frac{kM}{r^3}\, [\bv \times \br]\,\s
\,.
\end{equation}

\subsection{Spin-Orbit Interaction. Schwarzschild Field}

We consider now the spin precession in the Schwarzschild field
beyond the weak-field approximation (though neglecting the spin
influence on the trajectory). The 3-dimensional components of the
Schwarzschild metric can be conveniently written as
\begin{equation}\label{msch}
g_{mn}=\,-\,\left(\delta_{mn}-\,{r_m r_n \over r^2}\right)\,
-\,{r_m r_n \over r^2}\,{1 \over 1-r_g/r}\,
=\,-\,\delta_{mn}^{\perp}-n_m n_n\,{1 \over 1-r_g/r}\,.
\end{equation}
Nonvanishing tetrads are chosen as follows:
\begin{equation}\label{tsch}
e^{(0)}_0=\,\sqrt{1-r_g/r};\quad e^{(k)}_m
=\,\delta_{km}^{\perp}+n_k n_m\,{1 \over \sqrt{1-r_g/r}}\,;
\end{equation}
in this subsection the tetrad indices are singled out by brackets.
Now the nonvanishing Ricci coefficients (here their last indices
are world ones) are
\begin{equation}\label{rsch}
\gamma_{(0)(i)0}=\,-\frac{kM}{r^3}r_i\,;\quad
\gamma_{(i)(j)k}=\,{1-\sqrt{1-r_g/r} \over r^2}
\,(\delta_{jk}r_i-\delta_{ik}r_j)\,.
\end{equation}
At last, the precession frequency in this case is
\begin{equation}\label{osch}
\vom=\,-\,\bL\,{r_g \over 2 mr^3}\, \left\{{2 \over
u^0+u^0\sqrt{1-r_g/r}} +\,{1 \over 1+u^0\sqrt{1-r_g/r}}\right\}.
\end{equation}
Here $m$ and $\bL$ are the particle mass and orbital angular
momentum, respectively;
\[ u^0=\,{dt \over d\tau}\,=\,\left\{1-r_g/r -
({\bf n}\bv)^2(1-r_g/r)^{-1} - (\bv^{\perp})^2\right\}^{-1/2}. \]

Rather cumbersome general expression (\ref{osch}) simplifies
for a circular orbit. Here
\[ u^0=\,\left(1-\,{3kM \over r}\right)^{-1/2};\quad
L=mr\,\left({kM\over r}\right)^{1/2} \,\left(1-\,{3kM \over
r}\right)^{-1/2},\] so that
\begin{equation}\label{cosch}
\omega=\,{(kM)^{1/2}\over r^{3/2}} \,\left[1-\left(1-\,{3kM \over
r}\right)^{1/2}\right].
\end{equation}

The general case of spin precession in the Schwarzschild field was
considered in~\cite{ap}. Our expression (\ref{cosch})
agrees with the corresponding result of~\cite{ap} (therein the
precession is considered with respect to the proper time $\tau$,
but not with respect to $t$).

\section{Covariant and Noncovariant Description\\ of Spin-Dependent
Forces}

\subsection{Problem with Covariant Formalism. Simple Example}

The difficulty with the covariant description of spin-dependent
forces arises already for the electromagnetic interaction. To see
what we are talking about, let us come back to
Lagrangian~(\ref{la}). In the $c^{-2}$ approximation it generates
the spin-dependent force
\begin{equation}\label{thac}
f_m =\,\,\frac{eg}{2m}\s\B,_m+\,\frac{e(g-1)}{2m}
\,\left(\frac{d}{dt}[\E \times \s\,]_{,m} -\, \s [\bv \times
\E,_m\,]\right),
\end{equation}
acting on the particle (here and below in this subsection a
subscript with a comma denotes a partial derivative).

Let us try to construct a covariant expression for the
spin-dependent force acting on the particle, which would reproduce
in the same $c^{-2}$ approximation force~(\ref{thac}). Such
covariant correction $f^{\mu}$ to the Lorentz force
$eF^{\mu\nu}u_{\nu}$ should be linear in the tensor of spin
$S_{\mu\nu}$ and in the gradient of the tensor of electromagnetic
field $F_{\mu\nu,\lambda}\,$, it may depend also on the 4-velocity
$u^\mu$. Since $u^{\mu}u_{\mu}=1$, this correction must satisfy
the condition $u_{\mu}f^{\mu}=0$. From the mentioned tensors one
can construct only two independent structures meeting the last
condition. The first one,
\begin{equation}\label{co1}
\eta^{\mu\kappa}F_{\nu\lambda,\kappa}S^{\nu\lambda}\,-\,
F_{\lambda\nu,\kappa}u^{\kappa}S^{\lambda\nu}u^{\mu},
\end{equation}
reduces in the $c^{-2}$ approximation to
\begin{equation}\label{co11}
2\s(\B,_m-\,[\bv \times \E,_m]),
\end{equation}
and the second one,
\begin{equation}\label{co2}
u^{\lambda}F_{\lambda\nu,\kappa}u^{\kappa}S^{\nu\mu},
\end{equation}
reduces to
\begin{equation}\label{co21}
\frac{d}{dt}[\s \times \E],_m\,.
\end{equation}
Let us note that possible structures with the contraction
$F_{\nu\kappa,\lambda}S^{\kappa\lambda}$ reduce to (\ref{co1})
and (\ref{co2}), due to the Maxwell equations and the
antisymmetry of $S_{\kappa\lambda}$.

Certainly, no linear combination of (\ref{co11}) and (\ref{co21})
can reproduce the correct expression (\ref{thac}) for the
spin-dependent force.

But why is it that the correct (in the $c^{-2}$ approximation)
formula (\ref{thac}) cannot be obtained from a covariant
expression for the force? Obviously, one can easily reproduce by
a linear combination of (\ref{co11}) and (\ref{co21}) those terms
in (\ref{thac}) which are proportional to $g$. In other words,
there is no problem to present in a covariant form the terms
which describe, so to say, direct interaction of a magnetic
moment with external fields. It is the terms in (\ref{thac})
independent of $g$ and corresponding to the Thomas precession,
which cannot be written covariantly.

Of course, the noncovariance of equations by itself does not mean that
physical observables have wrong transformation properties. It is
sufficient to recall in this connection electrodynamics in the
Coulomb gauge.

\subsection{What Is the Coordinate of Spinning Particle?}

The covariant formalism can be reconciled with the correct results
if the coordinate $\x$ entering the covariant equation is related
to the usual one $\br$ in the $c^{-2}$ approximation as follows:
\begin{equation}\label{def}
\x\,=\,\br+\,\frac{1}{2 m}\s\times\bv.
\end{equation}
The generalization of this substitution to the case of arbitrary
velocities is~\cite{hei}
\begin{equation}\label{defh}
\x\,=\,\br+\,\frac{\gamma}{m(\gamma+1)}\s\times\bv,\;\;\; \gamma
=\,{1 \over \sqrt{1-v^2}}.
\end{equation}
Obviously, after this velocity-dependent substitution, the
Lagrangian depends explicitly on the acceleration which in general
results in spurious, nonphysical solutions.

Since relations (\ref{def}), (\ref{defh}) are valid for a free
spinning particle as well, their origin can be elucidated with a
simple example of a free particle of spin $1/2$. Here, instead of
the Dirac representation with the Hamiltonian of the standard form
$$H\,=\,\al \p\, + \,\beta\,m\,,$$
it is convenient to use the Foldy-Wouthuysen (FW) representation.
In it the Hamiltonian is
\[
H_{FW}=\beta \ep_{\bf p}, \quad \ep_{\bf p}=\sqrt{\p^2+m^2},
\]
and the 4-component wave functions $\psi_{\pm}$ of the states of
positive and negative energies reduce in fact to the 2-component
spinors $\phi_{\pm}$:
\[ \psi_+ = \left({\;\phi_+
\atop 0}\right), \quad \psi_- = \left({0 \atop \;\phi_-}\right)\,.
\]
Obviously, in this representation the operator of coordinate
$\hat{\br}$ defined by the usual relation
\beq\label{rrr}
\hat{\br}\psi(\br)=\br\psi(\br)\,,
\eeq
is just $\br$.

The transition from the exact Dirac equation in an external field
to its approximate form containing only the first-order correction
in $c^{-2}$, is performed just by means of the FW transformation.
Thus, in the resulting $c^{-2}$ Hamiltonian the coordinate of a
spinning electron is the same $\br$ as in the completely
nonrelativistic case \footnote{This is why nobody makes
substitution (\ref{def}) in the Coulomb potential when treating
the spin-orbit interaction in the hydrogen atom.}.

One more limiting case, which is of a special interest to us, is
a classical spinning particle. Such a particle is in fact a
well-localized wave packet constructed from positive-energy
states, i.e., it is naturally described in the FW representation.
Therefore, it is just $\br$ which it is natural to consider as the
coordinate of a relativistic spinning particle.

A certain subtlety here is that in the Dirac representation the
operator $\hat{\br}$ is nondiagonal. However, the operator
equations of motion certainly have the same form both in the
Dirac and Foldy-Wouthuysen representations. Correspondingly, the
semiclassical approximation to both is the same. In particular,
the time derivatives in the left-hand-side of classical equations
of motion are taken of the same coordinate $\br$, which serves as
an argument of the fields in the right-hand-side of these
equations.

As to the covariant operator $\hat{\x}$, it has the simplest form
in the Dirac representation: \beq\label{covar}
\hat{\x}_D=\,\sqrt{\frac{\ep}{m}}\beta\br_D\sqrt{\frac{\ep}{m}},
\eeq where $\hat{\br}_D $ is the operator acting on the wave
function in the Dirac representation according to the rule
(\ref{rrr}). The covariance of the matrix element $\psi^{\dagger}
\hat{\x} \psi$ is obvious: the matrix $\beta$ transforms
$\psi^{\dagger}$ into $\bar{\psi}$, and the factors
$\sqrt{{\ep}/{m}}$ are needed for the covariant normalization of
the wave functions.

Let us rewrite the operator $\hat{\x}$ in the FW representation.
The matrix $U$ of the FW transformation is \beq U=\frac{m+ \ep
-\beta \val \p}{\sqrt{2 \ep (m+ \ep )}}. \eeq The calculation,
which is conveniently performed in the momentum representation
where $\br_D=\,i\Na_{\bf p}$, results in the following
expression: \beq\label{rec} \hat{\x}_{FW}=\,U^{\dagger}
\hat{\x}_D U=\beta \left(\br + \frac{1}{m(m+\ep)} \s \times \p
\right) - \frac{1}{2m} [ (\val \p) \hat{\br} + \hat{\br} (\val
\p) ]. \eeq Here
\begin{equation}
 \s={1 \over 2} \, \left(
                            \begin{array}{rr}
                                \si & 0  \\
                                0   & \si \\
                             \end{array}
                       \right)
      \end{equation}
is the relativistic operator of spin. Let us note that different
components of the relativistic coordinate operator (\ref{rec}) do
not commute. If we confine to the space of the positive-energy
states, then we can put in (\ref{rec}) $\beta=1$ and drop the
terms with $\val$. In this way we arrive at expression
(\ref{defh}).

\subsection{Back to Gravitational Interaction}

Thus, there is however a serious problem with the covariant formulation
of the equations of motion of a spinning relativistic particle.
This equation, for instance, in the gravitational field,
\beq\label{pap}
{D \over D\tau}\left(m u_{\mu} - S_{\mu\nu}{D u^{\nu} \over D\tau}
\right) = -{1 \over 2} R_{\mu\nu\rho\sigma}
u^{\nu}S^{\rho\sigma}\,,
\eeq
contains the third time-derivative. As long as the term
\[ - {D \over D\tau}\left(S_{\mu\nu}{D u^{\nu} \over D\tau} \right), \]
with the third time-derivative, is treated perturbatively, no
special problem arises with it by itself. However, the inherent shortcoming
of equation~(\ref{pap}) (pointed out above for the corresponding electromagnetic
equation in the $c^{-2}$ approximation) is that beyond the perturbation theory it
evidently has spurious, nonphysical solutions.

But let us confine in (\ref{pap}) to the leading in spin
approximation:
\beq\label{pap1}
{D u_{\mu} \over D\tau} = -{1 \over 2m} R_{\mu\nu\rho\sigma}
u^{\nu}S^{\rho\sigma}\,.
\eeq
We note that the right-hand-side of this equation is the only
covariant structure possible here (up to a numerical factor). For
the nonrelativistic motion in the gravitational field created by a
mass $M$, equation (\ref{pap1}) reduces to
\beq \label{pap2}
\ddot{\x}=-\,{kM \over x^3}\,\x + 3\,{kM \over m x^3}\left[\bv
\times \s - ({\bf n} \bv){\bf n} \times \s -2{\bf n} ({\bf n}
[\bv\times \s])\right], \quad {\bf n}={\x \over x}\,.
\eeq
However, the equation of motion that follows from the Lagrangian
with potential (\ref{uls}) is somewhat different:
\beq \label{core}
\ddot{\br}=-\,{kM \over r^3}\,\br + 3\,{kM \over m r^3}\left[\bv
\times \s - \frac{3}{2}\,({\bf n} \bv){\bf n} \times \s
-\frac{3}{2}{\bf n} ({\bf n} [\bv\times \s])\right], \quad {\bf
n}={\br \over r}\,.
\eeq
The reason of the disagreement between (\ref{core}) and
(\ref{pap2}) is that they refer to the coordinates $\br$ and $\x$
defined differently (see (\ref{def})).

This discrepancy was pointed out long ago in~\cite{{bar}} where
the classical result (\ref{uls}) was derived from the scattering
amplitude for the Dirac particle in the gravitational field.

Quite typical explanation of the discrepancy was: ``The quantum
field theory of the spin-1/2 particle from which the classical
result was derived does not have any spin supplementary condition
($u^{\mu}S_{\mu}=0$ or $u^{\mu}S_{\mu\nu}=0$). This is because
field theories deal with point particles and not with extended
bodies.'' In fact, first of all, spin in the Dirac theory
certainly satisfies the mentioned constraint (in the sense of
expectation values). On the other hand, is the proton in a
gravitational field a point particle or an extended body? The
deuteron? The uranium nucleus? Obviously, an extended body can be
treated as a point particle, as long as we do not go into details
of its structure and as long as we do not consider its internal
excitations. This point should be emphasized since up to now one
may hear utterances similar to the above one on ``point particles
and extended bodies'', even from some well-known theorists.


\begin{thebibliography}{99}

\bibitem{mm} M. Mathisson, Acta Phys. Polon. \textbf{6}, 163
(1937).
\bibitem{pa} A. Papapetrou, Proc. Roy. Soc. London A
\textbf{209}, 248 (1951).
\bibitem{pau} W. Pauli, Collected Scientific Papers,
ed. by R. Kronig, V.F. Weisskopf\\ (John Wiley and Sons, 1964), v.
2, p. 544.
\bibitem{ros} T.O. Niinikoski, R. Rosmanith, Nucl. Instr. Meth. A
\textbf{225}, 460 (1987).
\bibitem{hei} K. Heinemann, DESY report DESY 96-229;\\
E-print archive physics/9611001.
\bibitem{hath} K.S. Thorne, J.B. Hartle, Phys. Rev. D
\textbf{31}, 1815 (1985)
\bibitem{kww} L.E. Kidder, C.M. Will, A.G. Wiseman,
Phys. Rev. D \textbf{47}, R4183 (1993)
\bibitem{blda} L. Blanchet, T. Damour, B.R. Iyer, C.M. Will,
A.G. Wiseman,\\ Phys. Rev. Lett. \textbf{74}, 3515 (1995)
\bibitem{cho} H.T. Cho, Class. Quantum Grav. \textbf{15},
2465, (1998);\\ E-print archive gr-qc/9703071
\bibitem{khp} I.B. Khriplovich, A.A. Pomeransky,
Phys. Lett. A \textbf{216}, 7 (1996);\\ E-print archive
gr-qc/9602004.
\bibitem{khp1} A.A. Pomeransky, I.B. Khriplovich, Sov. Phys. JETP {\bf 86}, 839
(1998);\\ E-print archive gr-qc/9710098.
\bibitem{fr} J. Frenkel, Z. Phys. \textbf{37}, 243 (1926).
\bibitem{th} L.H. Thomas, Phil. Mag. \textbf{3}, 1 (1927).
\bibitem{bmt} V. Bargman, L. Michel, V. Telegdi,
Phys. Rev. Lett. \textbf{2}, 435 (1959).
\bibitem{th1} L.H. Thomas, Nature \textbf{117}, 514 (1926).
\bibitem{dk}  Ya.S. Derbenev, A.M. Kondratenko, Sov. Phys. JETP \textbf{37}, 968 (1973).
\bibitem{dau} L.D. Landau, unpublished.
\bibitem{fok} A.D. Fokker, Kon. Akad. Weten. Amsterdam. Proc.
\textbf{23}, 729 (1921).
\bibitem{ap} T.A. Apostolatos, Class. Quantum Grav. \textbf{13},
              799 (1996).
\bibitem{bar} B.M. Barker, R.F. O'Connell, Gen. Rel. Grav.
\textbf{5}, 539 (1974).

\end{thebibliography}
\end{document}